\documentclass[10pt,conference]{IEEEtran}

\IEEEoverridecommandlockouts
\usepackage{cite}
\usepackage{amsmath,amssymb,amsfonts}
\usepackage{multirow}
\usepackage{graphicx}
\usepackage{textcomp}
\usepackage{xcolor}
\usepackage[inkscapearea=page]{svg}
\usepackage{adjustbox}
\usepackage{algorithm}
\usepackage{tikz}
\usepackage{algpseudocode}
\usepackage{caption}

\def\BibTeX{{\rm B\kern-.05em{\sc i\kern-.025em b}\kern-.08em
    T\kern-.1667em\lower.7ex\hbox{E}\kern-.125emX}}
\begin{document}

\title{Synchronization Control-Plane Protocol for Quantum Link Layer}

\author{\IEEEauthorblockN{Brandon Ru, Winston K.G. Seah, Alvin C. Valera}
\IEEEauthorblockA{\textit{School of Engineering and Computer Science} \\
\textit{Victoria University of Wellington }\\
Wellington 6140, New Zealand\\
Email: \{rubran, winston.seah, alvin.valera\}@ecs.vuw.ac.nz}
}

\maketitle

\begin{abstract}
Heralded entanglement generation between nodes of a future quantum internet is a fundamental operation that unlocks the potential for quantum communication. In this paper, we propose a decentralized synchronization protocol that operates at the classical control-plane of the link layer, to navigate the coordination challenges of generating heralded entanglement across few-qubit quantum network nodes. Additionally, with quantum network simulations using NetSquid, we show that our protocol achieves lower entanglement request latencies than a naive distributed queue approach. We observe a sixfold reduction in average request latency growth as the number of quantum network links increases. The Eventual Synchronization Protocol (ESP) allows nodes to coordinate on heralded entanglement generation in a scalable manner within multi-peer quantum networks. To the best of our knowledge, this is the first decentralized synchronization protocol for managing heralded entanglement requests.

\end{abstract}


\section{Introduction}
Quantum networks harness quantum mechanical phenomena to achieve superior performance compared to classical networks and present opportunities for novel use cases. Fundamentally, they enable the distribution of quantum states over long distances, facilitating the development of quantum communication protocols like quantum key distribution (QKD) \cite{Bennett_2014}, which offers theoretically unbreakable encryption. Quantum networks will also enable distributed quantum computing, where complex problems are more efficiently solved than any single quantum computer could on its own \cite{caleffi2022distributed}.

 Central to the operation of quantum networks is the phenomenon of entanglement, where pairs or groups of particles become interconnected in such a way that the state of one particle instantaneously affects the state of the other, regardless of the distance separating them \cite{born_einstein_letters}. Entanglement is a fundamental resource of quantum networks that leverages correlations between particles to form a quantum link between nodes \cite{rfc9340}. It is a critical resource for quantum communication, enabling protocols like quantum teleportation, which allows the transfer of quantum states between distant locations without moving the physical particles themselves. 

However, quantum networks operate under a set of constraints that are markedly different from those of classical networks. Quantum bits, or qubits, are highly susceptible to decoherence and loss, requiring error correction and fault-tolerant mechanisms that are still in developmental stages. The no-cloning theorem, which states that it is impossible to create an identical copy of an arbitrary unknown quantum state, imposes fundamental limits on the way information can be duplicated and distributed across the network  and complicates the design of quantum communication protocols \cite{no_cloning}. 

Despite their advanced capabilities, quantum networks cannot operate entirely independently and must be integrated with classical networks. Classical infrastructure plays a crucial role in tasks such as clock synchronization, control of quantum devices, and error correction. Classical communication is also necessary for the post-processing steps in quantum key distribution and for coordinating entanglement distribution across the network. Therefore, the successful deployment of quantum networks will rely on a hybrid approach that combines both quantum and classical technologies.

\begin{figure}
    \centering
    \includegraphics[width=1\linewidth]{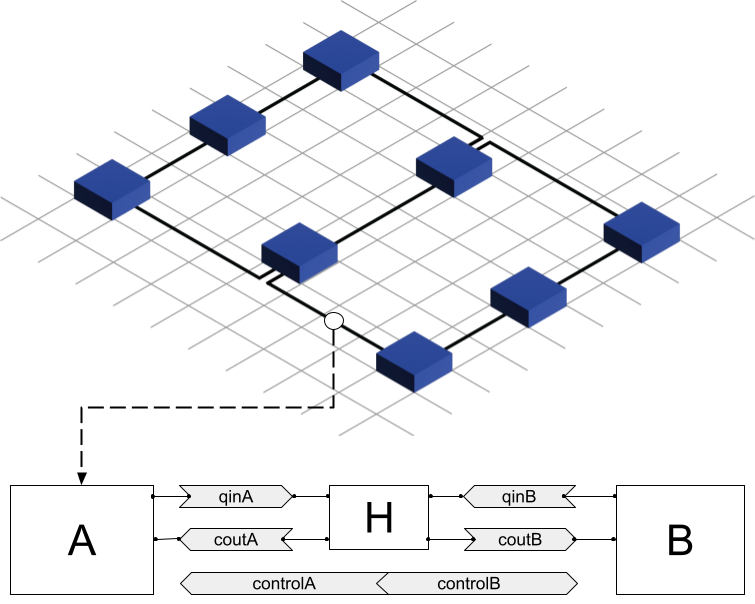}
    \caption{Example of an eight-node quantum network topology. Each pair of nodes generates heralded entanglement using the single-photon protocol, sending entangled qubits (photons) to the heralding station $H$ using inbound fiber-optic links. Entanglement success is probabilistic and heralded by outbound classical connections. Coordinating these entanglement requests involves an additional control plane topology.}
    \label{fig:headline}
\end{figure}

In this paper, we will build off the quantum link layer definition and protocol definition from Dahlberg et al. \cite{Dahlberg_2019}, and propose a control-plane protocol for synchronization. Although there is a mixed consensus regarding the definition of the link layer across other quantum network stack proposals \cite{Illiano_2022}\cite{Pirker_2019}\cite{Van_Meter_2009}, it is agreed that entanglement is a fundamental resource core to quantum networking.
 
 The proposed control-plane protocol is designed for synchronizing nodes in a multi-peer quantum network that wish to generate heralded entanglement. Fig.~\ref{fig:headline} illustrates an example of a quantum network topology the protocol could operate within. The protocol operates in a decentralized manner, without a central scheduler, and offers the opportunity for concurrent entanglement generation. Although other schemes exist for generating entanglement between separated nodes, recent research consisting of experimental trials in multi-peer quantum networks \cite{Liu2024} \cite{Hermans_2022} \cite{Pompili_2021} continue to use a heralded approach, which offers greater scalability and control to the end nodes. 
 
 This paper makes the following key contributions:

\begin{enumerate}
    \item \textbf{Decentralized Synchronization Protocol:} We introduce a control plane protocol designed for synchronizing heralded entanglement requests. Our quantum network simulations in multi-node topologies display a six-fold reduction in mean latency growth when compared to our baseline distributed queue protocol.

    \item \textbf{Simulated Benchmarks for DQP: } In our paper, we also extend the original DQP implementation in \cite{Dahlberg_2019} beyond a two-node network, and test it in a multi-peer quantum network to measure its performance.

    \item \textbf{Control Plane Definition:} We provide a clear definition of the control plane within the quantum link layer, emphasizing the interdependent relationship between classical and quantum networking required within the link layer.

\end{enumerate}

The rest of the paper is organized as follows. The background of quantum entanglement generation, the constraints of quantum networking, and related work are outlined. The design details of the Eventual Synchronization Protocol (ESP) are presented in Section V. The performance of ESP is outlined, evaluated and analysed in Section VI. Conclusions are in Section VII. 

\section{Theoretical Background}
\subsection{Quantum Bits (Qubits) and Entanglement}
Qubits are the foundational building blocks of quantum systems. They are two-state quantum systems, and represent the linear superposition of a ground state and excited state (0 or 1). Quantum states are sensitive to the environment and suffer from \textbf{decoherence}. Decoherence is the loss of coherence in a qubit, which is the fundamental property that allows the qubit to exist in a superposition of states. The causes of decoherence are specific to the physical implementation of the qubit, but vary depending on the material and temperature \cite{zarrabi2023case}. In general, interactions with the environment and the passage of time will introduce decoherence into a qubit.

Quantum entanglement is another fundamental phenomenon where two or more particles are linked in a way such that the quantum state of one particle cannot be independently described without considering the state of the others. As quantum entanglement is a phenomenon that does not depend upon distance or a physical connection, maximally entangled (Bell State) pairs of qubits can be used to create a quantum link between network nodes. This correlative phenomenon between Bell State pairs is also crucial to faciliate quantum teleportation, the transfer of a qubit along this entangled link.

\subsection{Heralded Entanglement}
Remote entanglement using emitted photons carrying a quantum state (single-photon protocol) is well-suited for entanglement generation between network nodes \cite{Hermans_2023}. Situated between the nodes is a \emph{heralding station H}, which is connected to the end nodes with an optical fibre link. Each end node, \emph{A} and \emph{B}, prepares a \emph{communication qubit} in the superposition state \(\sqrt{\alpha}|0\rangle + \sqrt{1-\alpha}|1\rangle\), where $\alpha$ is known as the \emph{bright state population} of the qubit. Excitation of the qubit in the state $|0\rangle$, the \emph{bright state}, produces a joint entangled quantum state consisting of the communication qubit and a photon. Then, the role of the heralding station, $H$, is to act as a beam splitter and interfere with the incoming photons from \emph{A} and \emph{B}. Depending on the measurement outcome of the photons at \emph{H}, the communication qubits at \emph{A} and \emph{B} are projected into one of the Bell State pairs, $|\Psi\rangle^+$ or $|\Psi\rangle^-$, thereby generating an entangled pair of qubits between adjacent network nodes. 

However, this heralding process is probabilistic and only succeeds with a small probability. This entanglement process is known as \emph{heralded entanglement generation} \cite{Bernien_2013}, and requires photons to only travel half the distance of a link and a stateless intermediate node. However, observe that \textbf{both} end nodes \emph{A} and \emph{B} are involved in this process, as they need to send their photons to \emph{H}. This suggests the need for a synchronization protocol to ensure end nodes coordinate before they attempt entanglement generation.

\subsection{Entanglement Fidelity}
Physical systems are not perfect and the entanglement generation is subject to noise. That is, a generated entanglement under perfect conditions would be a Bell State, but in reality is slightly different. \emph{Entanglement fidelity}, denoted by $F$, is a purely quantum metric that measures the quality of the generated entanglement, where $F \in [0, 1]$ and $F = 1$ is the ideal state. Entanglement fidelity is of importance as it directly affects the functionality of higher level services that consume these entanglements. For example, it is well known that there are strict bounds on the quantum bit error rate (QBER) in quantum key distribution (QKD) protocols, to ensure safety against eavesdropping. Lower entanglement fidelities used in QKD would result in a higher QBER, meaning entanglement fidelity is an important quality-of-service parameter that should be accounted for during entanglement generation. In practice, only entanglements where $F > 0.5$ contain useful amounts of entanglement for use \cite{Matsuo_2019}.

\begin{figure}
    \centering
    \includegraphics[width=1\linewidth]{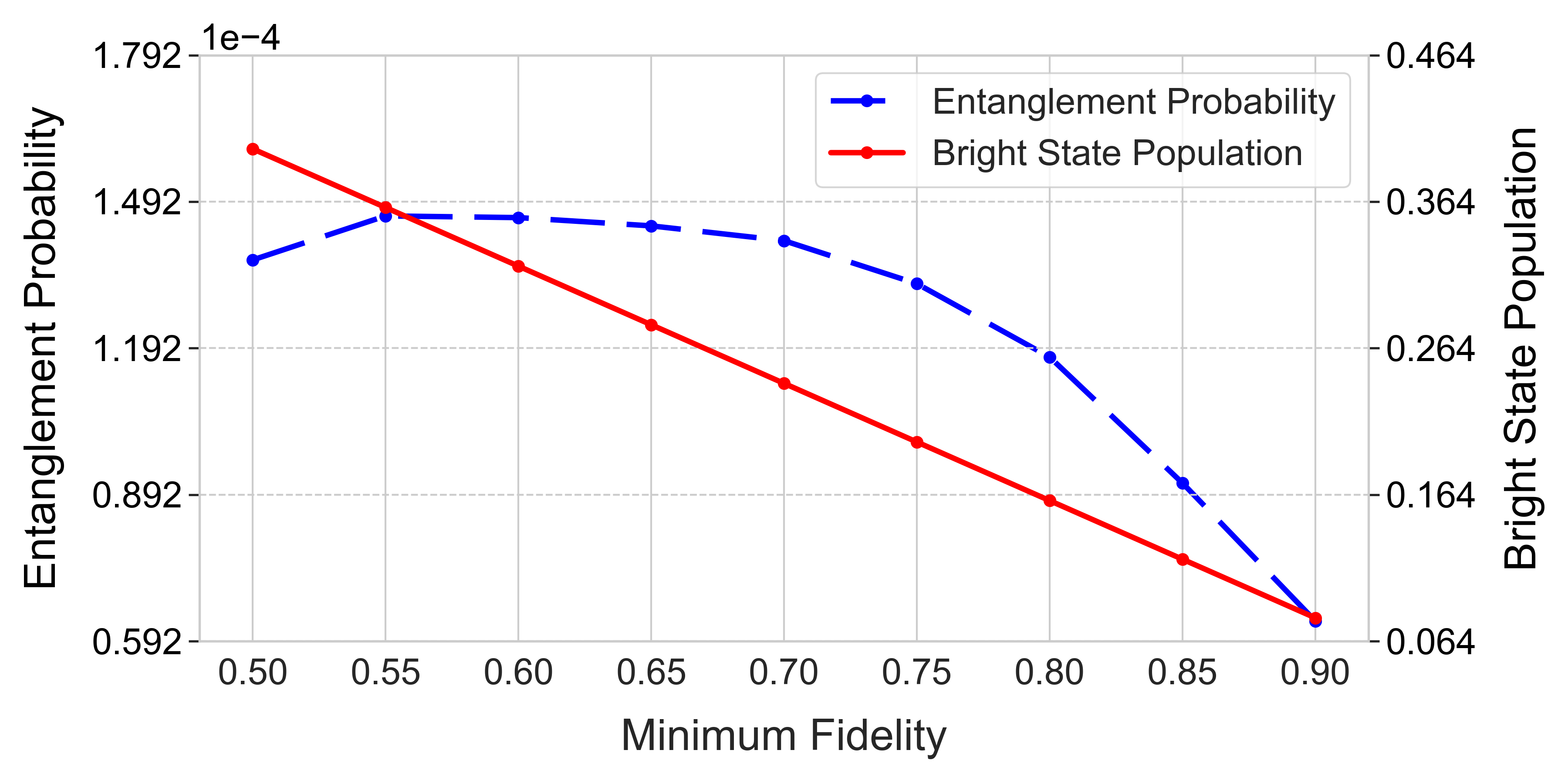}
    \caption{Relationship of the entanglement generation probability ($p_{\text{success}}$), minimum entanglement fidelity ($F_{\text{min}}$), and bright state population ($\alpha$). This relationship is used to model the physical layer entanglement generation process within our simulations. In this example, we obtain $p_{\text{success}}$ from a $t=120$ (seconds) simulation on $2\times2$ network topology.}
    \label{fig:p_vs_fidelity}
\end{figure}

The relationship between the bright state population of the single-photon heralded entanglement generation protocol, $\alpha$, and entanglement fidelity, $F$ is also of significance \cite{Hermans_2023}. Empirical simulated results are displayed in Fig.~\ref{fig:p_vs_fidelity} and they underpin the physical layer of our network simulations. Heralded entanglements generated with lower values of $\alpha$ for the initial communication qubit have a higher final entanglement fidelity and vice versa. However, there is also a tradeoff to consider as there is a positive relationship between $\alpha$ and $p_{success}$, the probability of a successful entanglement occurring at \emph{H} between the incoming photons. If each entanglement attempt requires time to execute, then there is a clear fidelity-time tradeoff, where higher fidelity entanglements require longer times to generate and vice versa.

\subsection{Quantum Link Layer}
Entanglement Generation Protocol (EGP) is an explicit definition of a protocol for robust entanglement generation \cite{Dahlberg_2019}. EGP manages the requests between adjacent nodes that wish to generate an entanglement, and is responsible for a wide range of tasks including scheduling, timekeeping, quantum memory management, and interfacing with a physical layer protocol.

One such physical layer protocol, the Midpoint Heralding Protocol (MHP), uses the single-photon scheme and is responsible for polling the higher EGP layer and attempting heralded entanglement between an adjacent node. Importantly, MHP contains no decision making elements and operates with discrete time steps at the nanosecond scale (i.e. the physical layer operates without state in an automated manner).

Hence, EGP relies on a synchronization protocol to ensure the remote node of the heralded entanglement attempt is aware of the request and it can succeed. This is because both end nodes of the entanglement must simultaneously send photons to the heralding station, \emph{H}, and is a general problem in heralded entanglement generation.

\section{Related Work}
Coordination is an inherent part of the quantum link layer and is achieved with various control plane protocols. The generation of heralded entanglement requires end nodes of the link to interact with the midpoint heralding station. For example, in EGP, a distributed queue protocol (DQP) is used to coordinate requests between layers \cite{Dahlberg_2019}. The distributed queue is implemented using a master-follower architecture, and requests are placed into the queue by the end nodes of a link using a two-way handshake. In contrast, our solution uses a decentralized architecture where all end nodes are equal, and communicate with one another.

Skrzypczyk \cite{skrzypczyk2021architecture} and Pompili et al. \cite{Pompili_2022} suggest a centralized approach is more appropriate, as a centralized scheduler can produce dynamic time division multiple access (TDMA) schedules for the network, given it has an overarching view of entanglement request details. This setup is similar to software-defined networking (SDN) schemes in classical networks, and allows the scheduler to account for quality-of-service requirements. However, even though the central scheduler can be implemented in a fault-tolerant distributed manner, it still poses as additional overhead to a local quantum network and raises scalability concerns. Our proposed protocol currently differs as it does not perform any explicit scheduling of requests, and operates with a first-in-first-out (FIFO) strategy.

Another control-plane protocol, RuleSet \cite{Matsuo_2019} tackles this coordination problem directly by introducing a declarative protocol that allows nodes to execute coordinated actions. Rules are distributed to nodes that need to perform shared actions, and consist of \emph{Conditions} (which are conditional statements) and \emph{Actions} which reserve and allocate resources to other nodes. However, it is not explicitly clear how this resource allocation and locking occurs, as there can be contention between nodes. The proposed solution hopes to offer a potential strategy for addressing this resource allocation, as it offers a means to synchronize requests.

In addition to explicit control plane protocol research, there are experimental trials to generate entanglement in a scalable manner within multi-peer quantum networks. For example, Hermans et al. \cite{Hermans_2022} performed qubit teleportation between non-adjacent nodes in a quantum network by first generating entanglement with a shared central node and performing entanglement swapping. Each node is optically connected in a chain, and utilize the single-photon protocol to generate entangled pairs using a communication qubit. 

Another recent endeavour has also shown promising experimental results on a metropolitan scale between three-nodes, maximally separated by 12.5 km, from Liu et al \cite{Liu2024}. The single-photon heralded entanglement scheme is also used, and each of the quantum end nodes are connected to a central server. The central server contains a multi-input two-output optical switch that allows the nodes to share a heralding station. However, hot-swapping between the pairs wishing to generate entanglement is required at the switch, and is akin to a time-sharing multiple access protocol.

At a classical level, there also exist protocols for managing contention and coordination. These include centralised lock managers which have a global view of requests and are able to provide desirable qualities such as the use of scheduling strategies \cite{lottery}\cite{mlfq}, and starvation and deadlock prevention techniques \cite{centralised_starvation}. Distributed protocols, such as the Ricart-Agrawala Algorithm \cite{ricart} and Lamport's Distributed Mutual Exclusion Protocol \cite{lamport2019time} also exist and whilst are effective for traditional distributed systems, do not fully transfer into the context of managing entanglement generation. In a quantum network, a node may receive multiple simultaneous requests to generate entanglement with different nodes, requiring a mechanism to defer and redirect attention to other nodes if a request is rejected, which is not adequately handled by these protocols.

\begin{figure*}[htbp]
  \centering
  \adjustbox{trim=0mm 5.75mm 0mm 6.75mm}{
  \includesvg[scale=0.45]{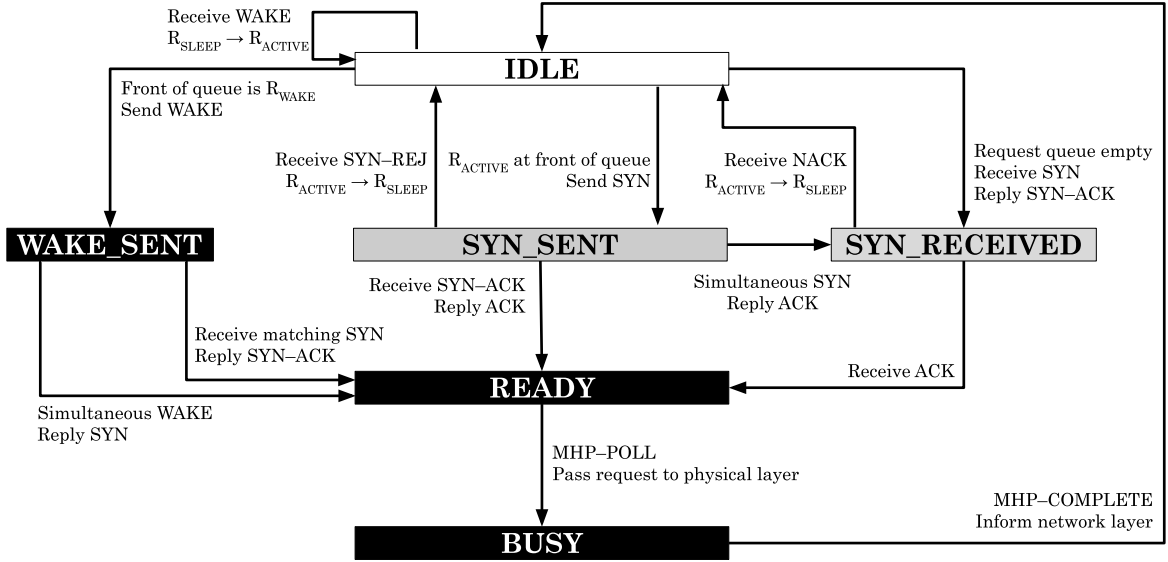}
  }
  \caption{ESP Finite State Machine Representation}\label{fig:esp}
\end{figure*}

\section{Problem Formulation}
A quantum network, represented by $G_Q = (V, E_Q)$, consists of quantum links, $E_Q$, between quantum nodes, $V$, forming the logical entanglement topology of the network. That is, $E_Q$ is equivalent to the set of possible entanglements that can be generated. During heralded entanglement, both nodes involved in the request must be aware of the request details and synchronized. This requires a separate control-plane network $G_C = (V', E_C)$, where each node must be able to make an entanglement request, $V \subseteq V'$,
but $E_C$ and $E_Q$ can be different.

In this model, we assume each quantum node can only participate in entanglement generation with exclusively one remote node, due to strict timing constraints from entanglement decoherence. That is, an invariant of the system is:

\begin{equation}
\sum_{j \in V, j \neq i} x_{ij} \leq 1  \: \: \: \forall i \in V
\end{equation}

\noindent where $x_{ij}$ indicates active entanglement generation between node $i$ and node $j$. The objective is to design a control plane protocol for efficient entanglement generation that operates on a suitable $G_C$ configuration, given the constraints of the quantum network $G_Q$.

Requests arrive sequentially from higher layers of the quantum network stack, denoted by the set $M = \{r_1, r_2, \dots, r_n\}$, where each request $r_k$ is defined as:

\begin{equation}
r_k = (i_k, j_k, f_k, p_k, t_k)
\end{equation}

where:
\begin{itemize}
    \item $i_k, j_k \in V$ are the initiator and remote nodes.
    \item $f_k$ denotes the fidelity requirement for the entanglement.
    \item $p_k$ is the number of entangled pairs requested.
    \item $t_k$ represents the time at which the request is made, with sequential arrivals such that $t_1 \leq t_2 \leq \dots \leq t_n$.
\end{itemize}

In this paper, our objective is to minimize the average request latency over $M$ within the network, through the design of an appropriate link-layer protocol that communicates using the control plane $G_C$. For comparison, we use the DQP protocol from \cite{Dahlberg_2019} as our baseline. In our implementation, DQP adopts a centralized scheduler, $S$, to manage the shared view of requests. Each node makes requests and obtains updates from $S$ and it follows that $G_C = (V \bigcup {S}, E_C)$ where $E_C = \{(i, S) \:\forall\: i \in V\}$. $S$ maintains a shared queue of all requests and updates, enabling nodes to synchronize properly before initiating entanglement generation on a given edge in $E_Q$. We choose this approach as our baseline because a centralized method has been proposed in \cite{skrzypczyk2021architecture, Pompili_2022}, and is a natural extension of the link-local implementation in \cite{Dahlberg_2019} to a multi-peer quantum network. It functions similarly to a centralized lock manager in a distributed system, offering a straightforward yet efficient means of managing coordination and contention.

\begin{table}[t]
\centering
\caption{Protocol Messages in ESP\vspace*{-6pt}}\label{tab:esp-messages}
\begin{tabular}{|l|l|}
\hline
Message  & Description                                                     \\ \hline
$SY\!N$    & Initial request to synchronize and acquire resources            \\ \hline
$SY\!N\!ACK$ & Request to synchronize can be met at remote node                \\ \hline
$SY\!N\!REJ$ & Request to synchronize cannot be met at remote node             \\ \hline
$ACK$    & Request is valid at initiator node, complete.         \\ \hline
$N\!ACK$   & Request is no longer valid at initiator node, reject. \\ \hline
$W\!AKE$   & Unblock request at remote node.                                 \\ \hline
\end{tabular}
\end{table}

\begin{table}[t]
\centering
\caption{Protocol States in ESP\vspace*{-6pt}}\label{tab:esp-states}
\begin{tabular}{|l|l|}
\hline
State         & Description                                               \\ \hline
\textbf{IDLE}
                & Node has not initiated any synchronization.     \\ \hline
\textbf{SYN\_SENT}
                & Node has sent out the initial $SY\!N$ message.                \\ \hline
\textbf{SYN\_RECEIVED}
                & Node has received a $SY\!N$ message and accepts it. \\ \hline
\textbf{WAKE\_SENT}
                & Node has sent out an unblocking signal.    \\ \hline
\textbf{READY}
                & Resource reserved. Synchronization complete.      \\ \hline
\textbf{BUSY}
                & Resource is in use.                              \\ \hline
\end{tabular}
\end{table}

\section{Eventual Synchronization Protocol (ESP)}
This section presents the Eventual Synchronization Protocol (ESP), a decentralized control-plane protocol for heralded entanglement. Fig.~\ref{fig:esp} illustrates the general control flow of ESP using a finite-state machine, while
Table~\ref{tab:esp-messages} lists the ESP messages, 
Table~\ref{tab:esp-states} details the states and transitions of the protocol's finite-state machine, and Table~\ref{tab:esp-implementation} outlines the relevant implementation details.

\begin{algorithm}
\caption{ESP Request Processor}\label{alg:cap}
\text{For each quantum node, running continuously:}
\begin{algorithmic}[1]
\If{$|Q_{ready}| > 0$}
    \State $R \gets Poll(Q_{ready})$
    \State \text{Execute $R$ at physical layer} \Comment{Non-blocking}
    \State On completion, Release(Lock)
\ElsIf{$|Q_{main}| > 0$}
    \State $P \gets Poll(Q_{main})$
    \If{$P$ is of type \textit{active}}
        \State Send $SY\!N\!$
        \State $S_{sent} \gets P$
    \ElsIf{$P$ is of type \textit{wake}}
        \State Send $W\!AKE$
        \State $S_{wake} \gets P$
    \EndIf
\Else
    \State Block until new request in $Q_{ready}$ or $Q_{main}$
\EndIf
\end{algorithmic}
\end{algorithm}

\subsection{Overview}
In ESP, each quantum node is treated as a resource that must be locked during the entanglement process. The control plane protocol should ensure that heralded entanglement generation only occurs between adjacent nodes which are properly synchronized. Specifically, for an edge $(i, j) \in E_Q$, node $i$ must acquire the \emph{Lock} on node $j$ to initiate entanglement. Otherwise, the entanglement process cannot proceed if both nodes in the edge $(i, j)$ are not participating. Entanglement requests are placed into a main queue $Q_{main}$ and are processed sequentially by Algorithm~\ref{alg:cap}.

\begin{figure}[h]
\centering
\label{fig:snapshot}
\begin{tikzpicture}[node distance={15mm}, thick, main/.style = {draw, circle}]
    \node[main, fill=lightgray] (1) {$A$}; 
    \node[main, fill=lightgray] (2) [right of=1] {$B$};
    \node[main, fill=lightgray] (3) [right of=2] {$C$};
    \node[main, fill=lightgray] (4) [below of=1] {$D$};
 \node[main, fill=purple!30] (5) [below of=2] {$E$};
    \node[main, fill=purple!30] (6) [below of=3] {$F$};
    \node[main, fill=purple!30] (7) [below of=4] {$G$};
    \node[main, fill=purple!30] (8) [below of=5] {$H$};
    \node[main, fill=lightgray] (9) [below of=6] {$I$};

    \draw[<-] (1) -- (2);
    \draw[<-] (1) -- (4);
    \draw[<-] (9) -- (6);

    \draw[dotted] (5) -- (6);
    \draw[dotted] (7) -- (8);
\end{tikzpicture}
\vspace*{0.2cm}
\caption{Example protocol state snapshot of a nine-node (3$\times$3) quantum network topology $G_Q$. E, F, G, and H are BUSY. C and I are IDLE. B and D are in SYN\_SENT. F has a queued $R_{wake}$ request for I.}\label{fig:9-node_Example}
\end{figure}
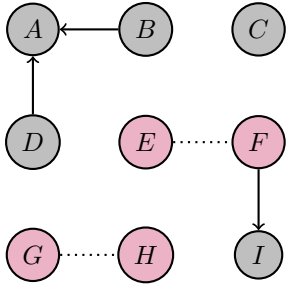

Instead of using a centralized scheduler $S$, in ESP, requests for entanglement generation are managed through the control plane network $G_C = (V, E_C)$ where $E_C = E_Q$. That is, the control plane network is equivalent to the logical quantum network topology. Synchronization requests are initiated with the SYN message, and put the initator into the SYN\_SENT state and if the remote node is IDLE, into the SYN\_RECEIVED state. In Fig.~\ref{fig:9-node_Example}, nodes B and D are in the state SYN\_SENT as they attempt to make entanglements with node A. If a node receives multiple SYN requests such as node A, it chooses the request that arrives first and enters the SYN\_RECEIVED state, and rejects the remaining options.

 A key concept in ESP is handling requests that cannot immediately acquire the remote node. For example, consider the request to generate entanglement along the edge $(i, j)$. If node $i$ makes the request and node $j$ is busy, then node $i$ puts the request to sleep and places it into $S_{sleep}$, essentially making the latter inactive. The protocol will exit the SYN\_SENT state and return to IDLE, allowing the node to process other pending requests without blocking. A given node is considered busy if it is has an acquired \emph{Lock}, that is, its state is in SYN\_RECEIVED, WAKE\_SENT, READY, or BUSY.

Nodes that reject requests with SYN\_REJ due to resource contention asynchronously signal the original request initiator when they become available again by queueing a wake-up request, $R_{wake}$ into the main queue $Q_{main}$. Upon processing of an $R_{wake}$ request, the node will enter the WAKE state. Then, upon receiving a WAKE message, a given node will remove a sleeping request from $S_{sleep}$ and place it back into $Q_{main}$. In Fig.~\ref{fig:9-node_Example}, nodes E and F are busy generating entanglements, and after node F finishes this process, it processes the $R_{wake}$ request to node I, which it must have rejected prior. Furthermore, the WAKE process cannot be interrupted by external SYN requests, to ensure  starvation from priority inversion cannot occur. That is, if node E decides to make another set of entanglements with node F while it is waking node I up, it will reject node E. This mechanism ensures that the original initiator is notified when previously unavailable resources are freed, thus achieving \emph{eventual synchronization} and an efficient management of entanglement requests.

\begin{table}[t]
\centering
\caption{Notation used in Protocol Design}\label{tab:esp-implementation}
\begin{tabular}{|l|l|}
\hline
Notation    & Description                                               \\ \hline
$S_{wake}$  & Set of requests where WAKE has been sent.                 \\ \hline
$S_{sent}$  & Set of requests where SYN has been sent.                  \\ \hline
$S_{sleep}$ & Set of requests that are sleeping.                        \\ \hline
$Q_{ready}$ & \begin{tabular}[c]{@{}l@{}}Queue of requests that are synchronized. \\ Ready for physical layer to process.\end{tabular} \\ \hline
$Q_{main}$ & \begin{tabular}[c]{@{}l@{}}Priority queue of requests that require synchronization. \\ Requests sorted in ascending order by arrival number.\end{tabular} \\ \hline
$Lock$           & Protected lock on physical layer.                    \\ \hline
\end{tabular}
\end{table}

Requests are assigned a local \emph{arrival number}, $A$, which is strictly sequentially increasing with each new request. This value is decided upon arrival of any type of request, including any wake requests, $R_{wake}$. Requests are maintained in $Q_{main}$, which is a priority queue such that requests with the lowest arrival numbers are processed first. When a sleeping request is reactivated and placed into $Q_{main}$, its arrival number remains unchanged. This ensures that requests that exit sleep are prioritized above newer ones to avoid ``starvation'' or indefinite waiting.

Each request also contains a minimum start time, $t_{min}$, as included in the original link-local protocol (DQP)~\cite{Dahlberg_2019}. $t_{min}$ accounts for the \emph{propagation time} of the link to prevent the entanglement attempt process from starting prematurely. This value is decided once the resource is reserved at the remote node, that is, when the state of the protocol enters a READY state at the remote node. This minimum start time is crucial because if the remote node initiates the entanglement process too early, it leads to unnecessary attempts that can ultimately impact the final fidelity of the produced entanglement.

The decentralized nature of ESP allows each node to manage its own state, eliminating the need for a centralized scheduler. This design scales effectively for larger networks with multiple quantum nodes and can be adapted to handle unstable network conditions through mechanisms such as timeouts, retry attempts, and expiration of stale requests. 

\section{Performance Evaluation}

In this section, the performance of ESP is compared against traditional network performance metrics as it operates at the control plane. Performance is compared against a baseline protocol, DQP, which consists of a centralized subnet-local scheduler.

\subsection{Performance Metrics}
\begin{enumerate}
    \item Request Latency, $L$ is defined as: 
    \begin{equation}L = NL_{finish} - NL_{start}
    \end{equation} 
    where $NL_{start}$ is the initial timestamp of requesting entanglement from the network layer, and $NL_{finish}$ is the final timestamp where all entangled pairs have been created.
    \item Request Jitter, $J$ is defined as the standard deviation of request latencies over a given time period:
    \begin{equation}
    J = \sqrt{\frac{1}{n} \sum_{i=1}^{n} (L_i - \overline{L})^2}
    \end{equation}
    where $L_i$ is the request latency of the $i$-th request, $\overline{L}$ is the mean request latency, and $n$ is the total number of requests in the given time period.
    \item Scaled Latency, $SL$ is defined as:
    \begin{equation}SL = \frac{L(R)}{|R|}\end{equation}
    where $|R|$ is the number of entanglements requested in $R$, therefore, this can be treated as a normalized latency.
   \item Scaled Network Throughput, $T$, is defined as:
    \begin{equation}
    T = \frac{\sum_{i=1}^{t} |E_i|}{E_Q \cdot t}
    \end{equation}
    where the numerator is the total number of entanglements that have been generated in the time period $t$, $E_Q$ is the number of edges in the topology $G_Q$, and the denominator is the product of the number of edges and the time period $t$.
    \item Busy Time, $B$ is given as: 
    \begin{equation}
    B = PL_{finish} - PL_{start}\end{equation} 
    where $PL_{start}$ is when $R$ is passed to the physical layer for entanglement generation, and $PL_{finish}$ is when the generation finishes. It does not include any queueing or transmission time.
    \item Queueing Time, $W$ is given as: 
    \begin{equation}W = PL_{start} - NL_{start}
    \end{equation} 
    where this time difference measures the conjunctive time $R$ has spent waiting in $Q_{main}$ and $Q_{ready}$, which begins when the request is created, $NL_{start}$, and ends when the request is passed to the physical layer, $PL_{start}$. 
\end{enumerate}

\subsection{Simulation Setup}
The performance of ESP is evaluated and compared using NetSquid (Python / C++), a discrete-event quantum network simulator developed by QuTech \cite{Coopmans_2021}.
Simplified versions of the original physical and link layer protocols, MHP and EGP,~\cite{Dahlberg_2019} are used to bootstrap our control-plane protocols for testing. Simulation settings are shown in Table~\ref{table:simulation-params}. We simulate entanglement generation in a quantum network, considering various parameters such as qubits per node, heralding station position, link length, photon detection window, and physical layer clock frequency. Requests for entanglement are generated using a bootstrap network protocol, representing an edge in the $G_Q$ topology. The simulation models the connections between nodes, quantum memory, processors, and the heralding station, ensuring accurate representation of the quantum network's physical and logical structures.

\begin{table}[h]
\centering
\caption{Simulation Parameters}\label{table:simulation-params}
\begin{tabular}{|p{4cm}|p{3cm}|}
\hline
\textbf{Parameter} & \textbf{Value} \\
\hline
Qubits per Node & 6 \\
\hline
Simulation Time & 120 seconds \\
\hline
Heralding Station Position & \( L / 2 \) \\
\hline
Link Length & 2 km \\
\hline
Photon Detection Window & 20 ns \\
\hline
Physical Layer Clock Frequency & Uniformly [450, 550] ns \\
\hline
Request Arrival Period & \( \lambda = 50 \) ms \\
\hline
Request Size & Uniformly [1, 6] \\
\hline
\end{tabular}
\end{table}

We employ a stochastic simulation approach. For each undirected edge in the topology, a request $R$ is made to generate an entanglement from one of the end nodes (randomly chosen). Once $R$ is satisfied, we sample from an exponential distribution to determine a backoff time period before making another request, $t_{b}$. This approach ensures that the entanglement requests follow a Poisson process, providing a realistic simulation of network behavior \cite{poisson}.
Finally, the size of each request, $|R|$ is uniformly sampled from $[1, 6]$, denoting the number of entanglements requested.

\subsection{ESP Evaluation}

\begin{figure}[t]
    \centering
    \includesvg[scale=0.6]{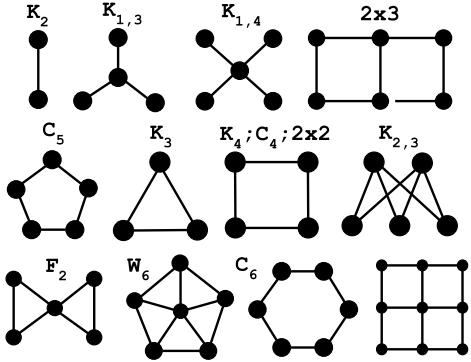}
    \caption{Network topologies used in the evaluation of ESP.}
    \label{fig:graph-topology}
\end{figure}

\begin{figure}[t]
    \centering
    \includegraphics[width=1\linewidth]{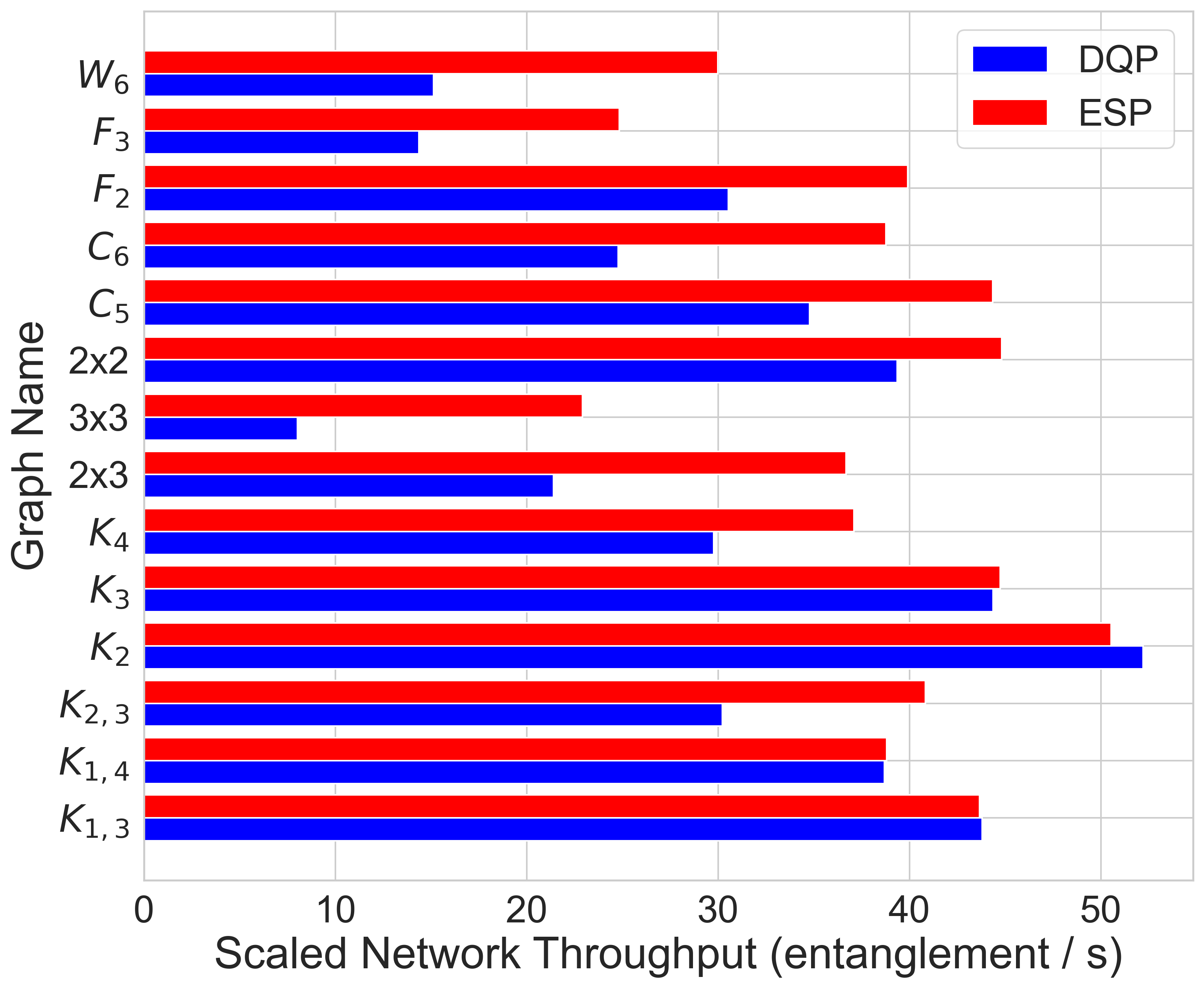}
     \caption{Scaled network throughput $T$ (entanglements / s) for each graph (Fidelity = 0.75), comparing DQP and ESP.}
    \label{fig:network-throughput}
\end{figure}

\begin{figure}[t]
    \centering
    \includegraphics[width=1\linewidth]{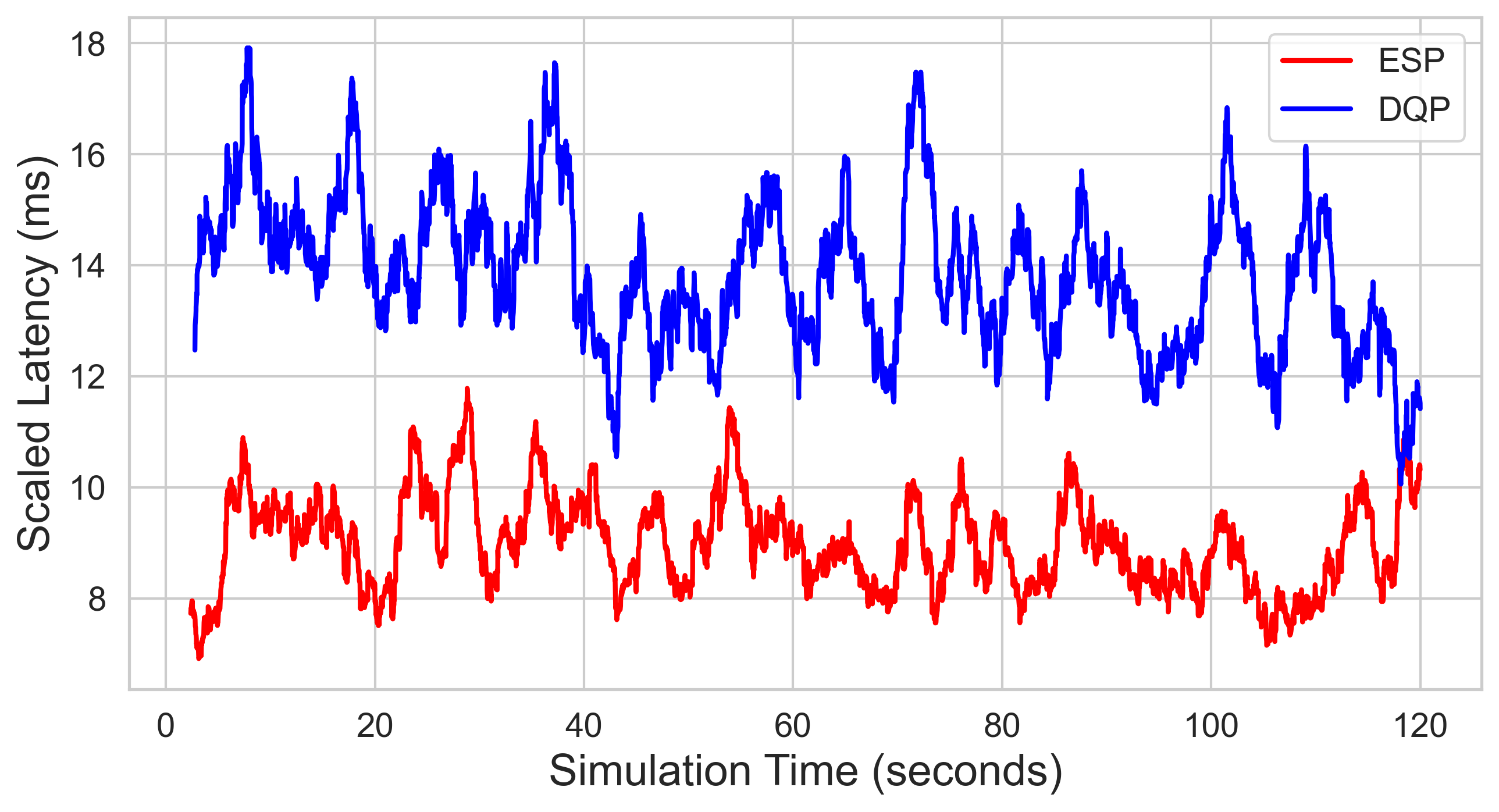}
    \caption{Time series of scaled latency $SL$ (ms) over a simulation period of \( t = 120 \) seconds for the \( 2 \times 2 \) \( G_Q \) topology.  A rolling average with \( k = 256 \) is applied due to the volume of requests processed.}
    \label{fig:scaled-latency-time-series}
\end{figure}

We perform our simulated evaluations on a range of network topologies $G_Q$~\cite{Gries1993}, including a set of $N{\times}M$ Manhattan grids (2$\times$2, 2$\times$3, 3$\times$3), complete graphs ($K_2$, $K_3$, $K_4$, $K_{2, 3}$), star graphs ($K_{1, 3}$, $K_{1, 4}$), friendship graphs ($F_2$, $F_3$), and wheel and cyclic graphs ($W_6$, $C_5$, $C_6$). These topologies are illustrated in Fig~\ref{fig:graph-topology}. For each network topology, we also evaluate the impact of entanglement fidelity $F$, and obtain results from the range [0.50, 0.90], with 0.05 increments. In this evaluation, we compare the performance of ESP  against the baseline DQP. Simulations are performed on a distributed computing cluster (ECS Grid) for total 368 simulated minutes ($\approx$ 4.42 billion physical layer cycles), across 1288 wall-clock hours. Table~\ref{table:latency_jitter} and Fig.~\ref{fig:network-throughput} show an overview of these results across the different topologies for $F=0.75$. An example time series graph for a $t=120$ network simulation on a 2x2 grid $G_Q$ is depicted in Fig.~\ref{fig:scaled-latency-time-series}.

\begin{table}[t]
\begin{center}
\begin{tabular}{|c|c|c|c|}
\hline
\textbf{Topology} & \textbf{Protocol} & \textbf{Mean Latency (ms)} & \textbf{Mean Jitter (ms)} \\
\hline
\multirow{2}{*}{$K_{1,3}$} & DQP & 29.48 & 20.28 \\
                           & ESP & 29.89 & 20.92 \\
\hline
\multirow{2}{*}{$K_{1,4}$} & DQP & 39.48 & 26.12 \\
                           & ESP & 39.50 & 25.79 \\
\hline
\multirow{2}{*}{$K_{2,3}$} & DQP & 66.31 & 34.57 \\
                           & ESP & 34.51 & 24.56 \\
\hline
\multirow{2}{*}{$K_{2}$} & DQP & 18.82 & 13.89 \\
                         & ESP & 18.75 & 13.47 \\
\hline
\multirow{2}{*}{$K_{3}$} & DQP & 29.31 & 20.32 \\
                         & ESP & 28.57 & 20.58 \\
\hline
\multirow{2}{*}{$K_{4}$} & DQP & 67.24 & 35.62 \\
                         & ESP & 44.80 & 30.57 \\
\hline
\multirow{2}{*}{2x2} & DQP & 39.19 & 25.50 \\
                              & ESP & 27.83 & 20.28 \\
\hline
\multirow{2}{*}{2x3} & DQP & 84.61 & 39.49 \\
                              & ESP & 33.03 & 24.32 \\
\hline
\multirow{2}{*}{3x3} & DQP & 180.11 & 48.86 \\
                            & ESP & 39.77 & 29.69 \\
\hline
\multirow{2}{*}{$C_5$} & DQP & 50.47 & 29.48 \\
                            & ESP & 27.67 & 19.78 \\
\hline
\multirow{2}{*}{$C_6$} & DQP & 65.91 & 33.95 \\
                            & ESP & 28.08 & 20.26 \\
\hline
\multirow{2}{*}{$F_2$} & DQP & 64.15 & 32.95 \\
                            & ESP & 38.11 & 28.04 \\
\hline
\multirow{2}{*}{$F_3$} & DQP & 119.40 & 43.39 \\
                            & ESP & 53.84 & 41.40 \\
\hline
\multirow{2}{*}{$W_6$} & DQP & 138.70 & 44.33 \\
                            & ESP & 54.34 & 38.97 \\
\hline
\end{tabular}
\end{center}
\caption{Comparison of ESP and DQP across various $G_Q$ topologies (Fidelity = 0.75), measured in milliseconds.} \label{table:latency_jitter}
\end{table}

\begin{figure}[t]
    \centering
    \includegraphics[width=1\linewidth, trim=0pt 12pt 0pt 0pt]{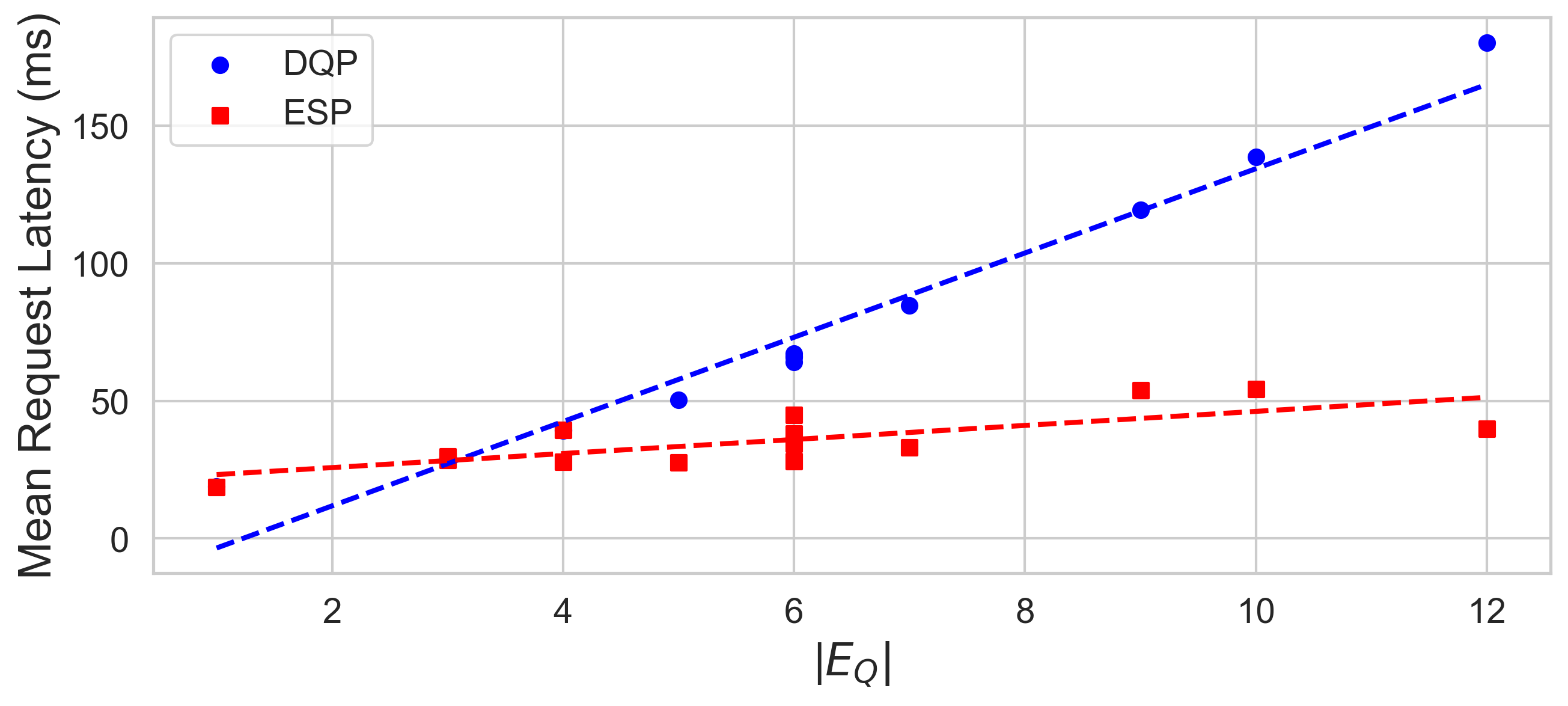}
    \caption{Relationship between the total number of edges $|E_Q|$ in $G_Q$ and the mean request latency $\overline{L}$, with dashed regression lines for each protocol (Fidelity = 0.75).}
    \label{fig:edges-vs-latency}
\end{figure}

\begin{figure}[t]
    \centering
    \includegraphics[width=1\linewidth, trim=0pt 12pt 0pt 8pt]{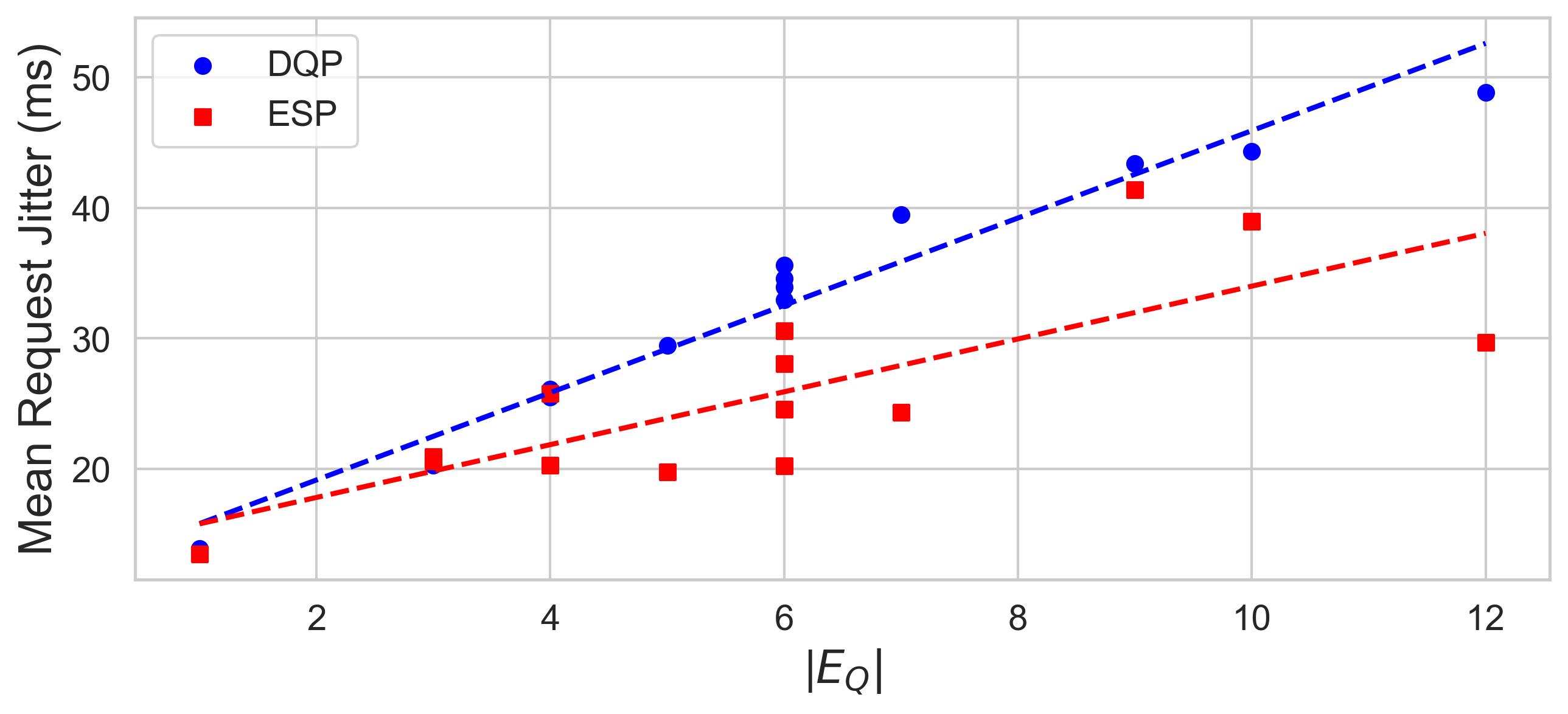}
    \caption{Relationship between the total number of edges $|E_Q|$ in $G_Q$ and the mean request jitter $\overline{J}$, with dashed regression lines for each protocol (Fidelity = 0.75).}
        \label{fi:edges-vs-jitter}
\end{figure}

\subsection{Impact of Network Size - $G_Q$} 
Fig.~\ref{fig:edges-vs-latency} illustrates how the average request latency $\overline{L}$ of ESP compares to that of DQP as the number of edges $|E_Q|$ in the network increases. As the network size grows, the total number of requests entering the system also increases, since the arrival rate $\lambda$ depends on each edge $(i, j) \in E_Q$. In DQP, the distributed queue processes requests only from the front of the queue and does not allow requests that could execute without contention to proceed. In contrast, ESP does not use a shared queue, which alleviates this bottleneck. Each node in ESP can independently execute requests if they are synchronized, allowing for concurrent entanglement generation across the overall network. The slope values for DQP and ESP are $m=15.32$ and $m=2.55$, respectively, with $R^2$ Pearson coefficients of $0.96$ and $0.54$. This indicates that ESP scales more efficiently than DQP, maintaining significantly reduced ($\sim\!6\times$ lower) latency values as the network expands.

Fig.~\ref{fi:edges-vs-jitter} shows the relationship between average request jitter $\overline{J}$ for ESP and DQP as the network size increases. Given the probabilistic nature of entanglement generation, some level of constant jitter is anticipated. We can see this through the intercept values for ESP and DQP,  $c=12.46$ and $c=13.77$ respectively, which are similar due to this inherent jitter.  The results indicate that while mean jitter increases with network size for both ESP and DQP, with slope values of $m=3.34$ and $2.02$ respectively, ESP exhibits a slightly smaller rate of increase. This is likely due to ESP's inherently lower mean request latency, which contributes to more stable performance and reduced jitter as the network expands. However, as both $m$ values are relatively similar and smaller in magnitude compared to $c$, we conclude the majority of the jitter still originates from the entanglement generation process and not the control plane protocol.

\subsection{Impact of Matching Number of Network - $\nu(G_Q)$}
\begin{figure}[t]
    \centering
    \includegraphics[width=1\linewidth]{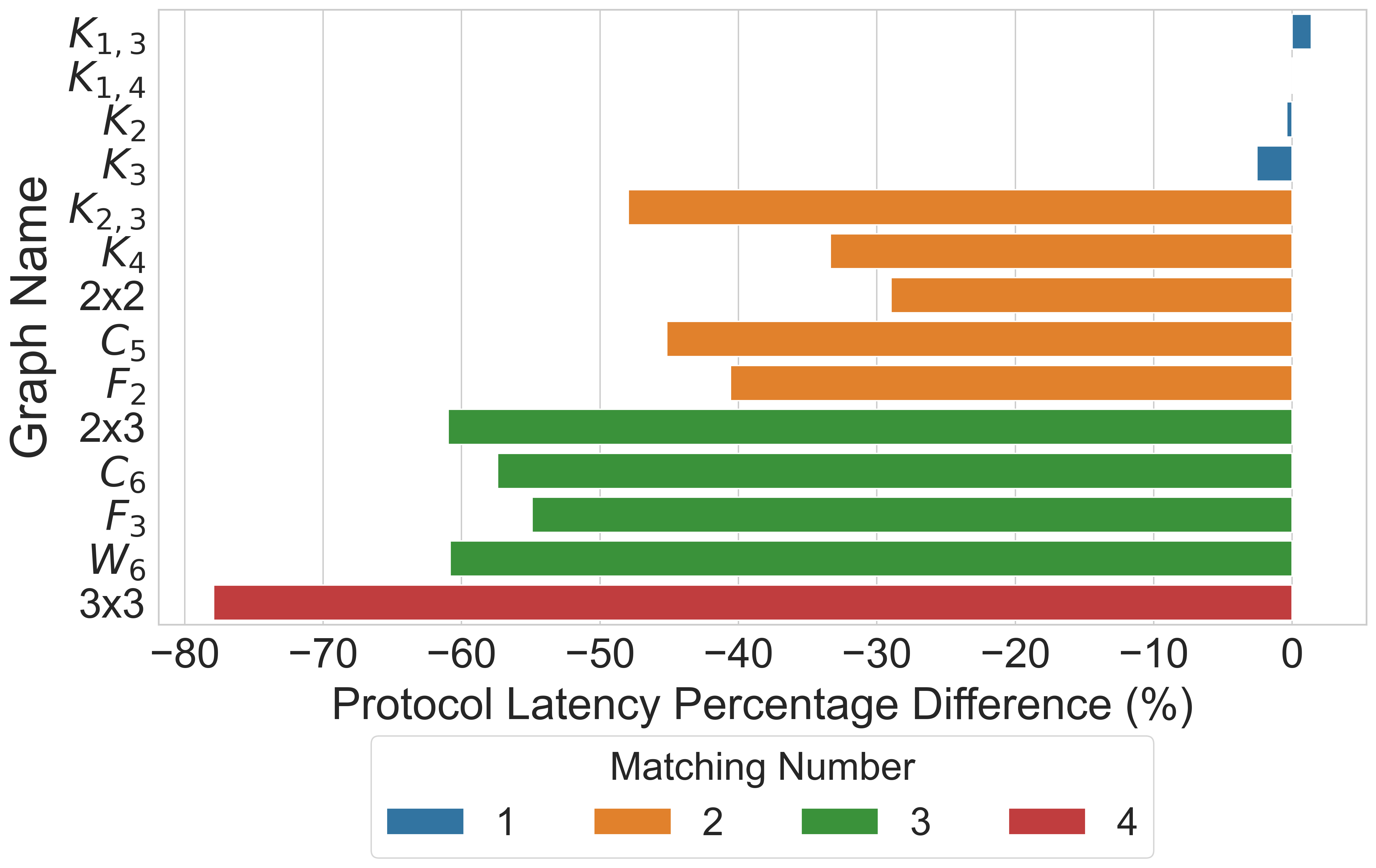}
    \caption{Illustrative results showing the percentage differences in mean request latency $\overline{L}$(ms) between ESP and DQP, categorized by the quantum network's matching number $\nu(G_Q)$.}
    \label{fig:percent-difference}
\end{figure}

In graph theory, a \emph{matching} is a set of edges in a graph where no two edges share a common vertex. For a quantum network, a matching represents a set of simultaneous requests that can be executed without conflicts. The \emph{matching number} of a graph, denoted $\nu(G_Q)$, is defined as the size of the largest possible matching in the graph $G_Q$. For example, in the graph shown in Fig.~\ref{fig:9-node_Example} ($3\times3$), the set of edges performing entanglement generation, \(\{(E, F), (G, H)\}\), represents a valid matching of $G_Q$, where \(\nu(G_Q) = 4\). As $\nu(G_Q)$ increases, ESP can potentially exploit this by handling more concurrent requests within the network. Fig.~\ref{fig:percent-difference} illustrates that ESP does not show substantial benefits in networks with $\nu(G_Q) = 1$ and may even suffer from performance degradation due to additional overhead ($K_2$, $K_3$, $K_{1, 3}$ $K_{1, 4}$). This behaviour is observed in smaller graph topologies and star graphs. However, for networks where $\nu(G_Q) > 1$, ESP demonstrates significant reductions in latency over DQP by efficiently allocating concurrent requests.

\subsection{Impact of Entanglement Fidelity $F$} 
\begin{figure}[t]
    \centering
    \includegraphics[width=1\linewidth]{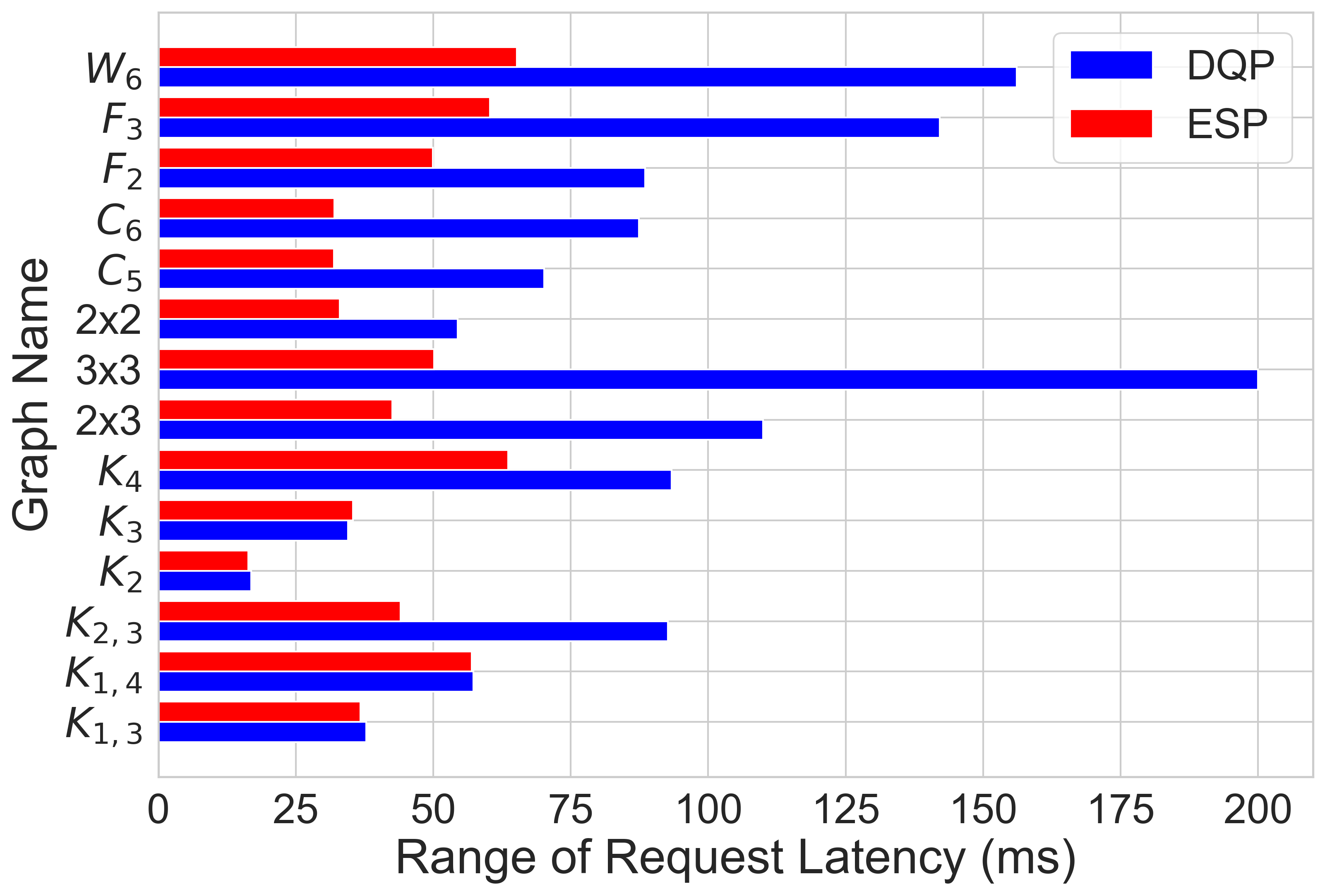}
    \caption{Bar chart illustrating the range of request latency (ms) from $F=[0.5,0.9]$ for each graph, comparing DQP and ESP.}
    \label{fig:range-request-latency}
\end{figure}

As shown in Fig.~\ref{fig:p_vs_fidelity}, higher entanglement fidelity of a request reduces $p_{success}$ at the physical layer, leading to an increase in busy time. This means each entanglement takes longer to generate, spreading a greater load through the network as queues take longer to process. Fig.~\ref{fig:range-request-latency} demonstrates this effect, illustrating the range of request latency between $F=0.5$ and $F=0.9$. In smaller networks, the impact on $\overline{L}$ is less noticeable due to minimal time spent queueing. For instance, in graphs $K_2$, $K_3$, and $K_{1,3}$, the ranges are similar between DQP and ESP. However, in larger graphs, the effect is significantly amplified, and increased entanglement fidelity results in larger increases in $\overline{L}$. Results show that ESP outperforms DQP in managing the effects of increased fidelity requests within the overall network by producing a tighter range of request latencies. 

\section{Conclusions and Future Work}

In this paper, we developed the Eventual Synchronization Protocol (ESP), designed for managing heralded entanglement requests in a multi-node quantum network. To the best of our knowledge, this is the first decentralized control-plane protocol for the quantum link layer. Our protocol operates in a distributed manner, enabling concurrent entanglement generation across the network. This approach results in a substantial increases in system throughput compared to our baseline protocol (DQP), which relies on a centralised scheduler. The advantages of our protocol are particularly notable in larger network topologies, where it effectively reduces the overall network impact associated with high-fidelity requests. We observe a sixfold reduction in average request latency growth as the number of quantum network links increases.

Future research will focus on improving the robustness of the protocol. This will involve investigating failure modes, addressing unstable network conditions, and validating certain distributed properties. Additionally, we plan to explore its potential as a resource reservation protocol, aiming to further optimize quantum network performance and the efficient management of entanglement resources.

\section*{Acknowledgment}
This work is supported by the Royal Society of New Zealand through a James Cook Research Fellowship.

\bibliographystyle{IEEEtran}
\bibliography{main}

\end{document}